\documentclass[11pt,aps,shownopacs,superscriptaddress,nofootinbib, preprint]{revtex4}
\usepackage{graphics}
\usepackage{graphicx}
\usepackage{epsfig}
\usepackage{psfrag}
\usepackage{color}

\usepackage{amsfonts}
\usepackage{amssymb}

\newcommand*{\be}{\begin{equation}}
\newcommand*{\ee}{\end{equation}}
\newcommand*{\bea}{\begin{eqnarray}}
\newcommand*{\eea}{\end{eqnarray}}

\providecommand*{\ger}{\stackrel{\scriptstyle >}{\scriptstyle \sim}}

\newcommand{\comment}[1]{}

\newcommand{\cref}[1]{Chapter~\ref{c.#1}}

\def\beq{\begin{equation}}
\def\eeq{\end{equation}}
\def\bea{\begin{eqnarray}}
\def\eea{\end{eqnarray}}
\def\ba{\begin{array}}
\def\ea{\end{array}}
\def\bi{\begin{itemize}}
\def\ei{\end{itemize}}
\def\be{\begin{enumerate}}
\def\ee{\end{enumerate}}
\def\bc{\begin{center}}
\def\ec{\end{center}}
\def\bt{\begin{table}}
\def\et{\end{table}}
\def\btb{\begin{tabular}}
\def\etb{\end{tabular}}

\def\lsim{\raise0.3ex\hbox{$\;<$\kern-0.75em\raise-1.1ex\hbox{$\sim\;$}}}
\def\gsim{\raise0.3ex\hbox{$\;>$\kern-0.75em\raise-1.1ex\hbox{$\sim\;$}}}

\begin{document}

\rightline{IISc/CHEP/13/07}
\rightline{August 2007}

\title{Results from MiniBooNE}
\author{Ranjan Laha}
\affiliation{Department of Physics, Indian Institute of
Science, Bangalore 560 012, India}
\author{Sudhir K. Vempati}
\affiliation{Centre for High Energy Physics, Indian Institute of
Science, Bangalore 560 012, India}

\begin{abstract}
The long awaited experimental results from MiniBooNE have recently been 
announced. This experiment tests whether neutrino oscillations
can occur at a higher mass squared difference $\sim1~\mbox{eV}^2$ 
compared to well established observations of solar and atmospheric 
neutrinos. The LSND experiment has previously claimed to have observed 
neutrino oscillations at $\Delta m^2 \sim 1 \mbox{eV}^2$, however the
results being controversial, required an independent confirmation. The
MiniBooNE results settle this controversy by observing null oscillations
at the said mass squared difference. These results have strong implications
on existence of sterile neutrinos, CPT violation and mass varying neutrinos.
We review the present status of neutrino masses and mixing in the light of 
this recent result. 
\end{abstract}
\maketitle

\section{Introduction }
In the Standard Model (SM) of elementary particles, there are three 
neutrinos, one for each flavour: electron type ($\nu_e$), muon-type ($\nu_\mu$)
and the tau-type ($\nu_\tau$). They do not have electric charge and 
participate only in weak interactions which are responsible for 
processes like nuclear $\beta$-decay etc. The original standard model,
not having enough knowledge of the other main physical attribute of the
neutrino, namely, their mass, has left them massless. However starting
from 1998, experimental measurements of neutrino oscillations have become
very robust implying neutrinos do have masses, however tiny. The neutrino
oscillation formula is given in terms of the mass squared differences and
the mixing angle parameters of the neutrinos. For the simplest case of 
two flavours, denoted by $a$ and $b$, oscillation probability is given by :
\beq
\label{prob}
P_{ab} = \mbox{sin}^2  2 \theta~ \mbox{sin}^2\left({ 1.27~ \Delta m^2_{ab} (\mbox{eV}^2) 
~L_\nu(\mbox{Meters}) \over E_\nu (\mbox{MeV}) }\right),
\eeq
with $\theta$ representing the mixing angle between the two flavours and $\Delta m^2_{ab} 
~=~m_{b}^2 - m_{a}^2 $,
the mass squared difference between them. $L_\nu$ and $E_\nu$ represent the distance
and the energy traversed by the neutrino respectively.  Within the standard picture 
of three neutrinos, which we elaborate below, there can be two independent mass 
squared differences responsible for the observed solar and atmospheric neutrino 
oscillations. 

One of the important challenges in this field was whether neutrino oscillations
would be observed not just in neutrinos produced in astrophysical processes, 
but, also in laboratory-like conditions, for example, neutrinos produced at
nuclear reactors or in particle accelerators. These experiments are of 
two types either Short-base line (SBL) or Long Base line (LBL), depending on 
the length neutrino traverses from the time of production to the time of 
detection.  One of the first claims for observation of neutrino oscillations
in laboratory was done by the Liquid Scintillation Neutrino Detector (LSND)
experiment conducted at the Los Alamos National Laboratory, USA.  However, 
this result soon ran in to controversy for various technical reasons\footnote{The LSND
evidence soon was termed as LSND anamoly as questions were raised regarding the 
accuracy about background estimates,etc. The LSND collaboration has responded to
most of the criticisms with elaborate checks. The evidence still persisted.} 
as well as for predicting existence of newer exotic particles
called sterile neutrinos. The oscillations observed  required a much larger 
mass difference compared to those required in solar and atmospheric 
oscillations and thus could only be explained by introducing a new neutrino 
which does not even participate in weak interactions and hence sterile. 

A second experiment called KARMEN failed to settle this controversy as it
could not probe the entire parameter space of the LSND experiment. The 
MiniBooNE was designed specifically to settle this controversial issue 
and prove/refute the simplest and popular explanation of the LSND result
\textit{i.e} the existence of a sterile neutrino at that mass range. 
This April, the MiniBooNE collaboration has announced its first results
after taking data for almost five years. Using statistically robust 
methods in their data analysis, they have found no positive signal for
neutrino oscillations at mass squared difference 
$\Delta m^2 ~\sim 1~ \mbox{eV}^2$. This result settles the 
LSND controversy which has dogged the particle physics community 
for over a decade. However caveats still do exist as we will explain
later.

In the present article, we report on this new experimental results and
comment on the implications the results would have on our understanding
of sterile neutrinos. The rest of the article is organised as
follows  : in the next section, we summarise the existing standard
picture of three neutrino oscillations. The summary is not necessarily
chronological in order, but we will give the dates wherever we can. 
In section 3, we elaborate on the LSND experimental results and their 
possible theoretical explanations. We also report on KARMEN's failure 
to contradict/validate the LSND experiment. In section 4, we report 
on the first results from MiniBooNE and their implications on particle 
physics scenarios. We close with some remarks on future directions. 

\section{The Standard Picture of three neutrino oscillations}

Neutrino oscillations were first proposed by Pontecorvo\cite{pontecorvo}
inspired by the observed neutral K-meson oscillations. The first experimental
indications of neutrino oscillations came from pioneering experiments of 
Raymond Davis Jr. (Nobel Laureate, 2002) measuring the neutrino flux
from the Sun. The Sun, as we know produces energy through nuclear fusion,
which can be summarised by the equation \cite{grimus}: 
\begin{equation}
4~p + 2~ e^{-} \rightarrow ^{4}He + 2~ \nu_{e} + Q,
\end{equation}
which shows four protons and two electrons fuse to form a Helium nucleus
giving out energy (Q = 26.73 million electron volts (MeV)) and two 
electron-type neutrinos ($\nu_e$). The expected number of $\nu_e$ coming
from the Sun to be observed at the earth can be computed using detailed
numerical computations, following the Standard Solar Model (SSM). However
observed number always fell short by about $50~\%$ compared to the expected
number giving rise to the so-called `Solar Neutrino Problem'\footnote{ This
problem persisted for over thirty years.}. The simplest solution proposed 
for the Solar Neutrino Problem was neutrino oscillations of Pontecorvo which 
are possible if neutrinos have tiny but non-zero masses. In such a case, 
the $\nu_e$ produced in the Sun, gets converted in to a $\nu_\mu $ or 
$\nu_\tau$ or a more exactly a linear combination of them while traversing 
the distance from the Sun to the detector placed on earth. It should be
noted that earlier detectors were sensitive only to $\nu_e$ \textit{i.e}
they could only detect $\nu_e$ but not $\nu_\mu$ and $\nu_\tau$ flavours. 
And hence, experiments could only validate that solar electron neutrinos 
do convert to $\nu_\mu$ and $\nu_\tau$ flavours conclusively only in late 2002. This 
was done by a combination of experiments at SNO (Sudbury Neutrino Observatory) 
in Canada, which was sensitive to all the three flavours, 
and at Super-Kamiokande detector located in Japan\cite{ananth}. 

However, there was still one more issue to be settled. This issue is 
concerned with the question how and where exactly the electron neutrinos
which are produced at the core of the Sun get converted to the other flavours 
while traversing the distance from the centre of the Sun to the Earth's
surface. In particular, taking into consideration the interaction of the 
neutrino with the dense matter of the Sun, another mechanism to convert 
$\nu_e$ to $\nu_{\mu(\tau)}$ called the MSW (Mikheyev, Smirnov and Wolfenstien) 
mechanism can happen other than the afore mentioned oscillations of the neutrino
in vacuum. It was thus important to identify exactly which mechanism was responsible
for the conversion of the electron neutrinos from the Sun as they reach the Earth.
This issue was recently settled by the experiment called KamLand which 
observed neutrino oscillations on the earth corresponding to the mass differences 
of the solar neutrinos. Finally, the data from all the experiments namely, 
KamLand, SNO, SuperKamiokande taken together points out to a large mixing 
MSW solution to the solar neutrino problem. The mass-squared difference 
and the mixing angle are determined to be\cite{garcia}: 
\begin{eqnarray}
\Delta m^{2}_{\mbox{solar}} &=& 7.9^{+0.27}_{-0.28} (^{+1.1}_{-0.89}) 
\times 10^{-5} 
\mbox{eV}^{2} \nonumber \\
\theta_{\mbox{solar}} &= &33.7 \pm 1.3 (^{+4.3}_{-3.5}) \deg,
\end{eqnarray}
where we have shown the errors bars in the 1$\sigma$ (3$\sigma$) range.

Atmospheric neutrinos have been discovered in  India and in South Africa in 
the 1960's as background for proton decay experiments. The origin of 
these neutrinos was traced to the interactions of cosmic rays with the 
atmospheric air molecules which led to the prediction for the ratio
 \beq 
{N_{\nu_{\mu}} + N_{\overline{\nu_{\mu}}} \over N_{ \nu_{e} }
+N_{\overline{\nu_{e}}}} \simeq 2 ,
\eeq 
where $N_{\nu_f}$ stands for the total number of the neutrinos corresponding
to the flavour $f$.  The bar on the top represents an anti-particle. 
This ratio is roughly expected to be `2' based on simple
analysis of Pion and Kaon decays. Detailed numerical simulations including
earth magnetic field effects also confirm this ratio to be close to `2'.
However, experiments using huge water Cerenkov neutrino detectors 
like IMB and Kamiokande observed a deviation from the above prediction, 
which can be best expressed in terms of a double ratio given by 
\beq
R = {(N_{\nu_\mu} / N_{\nu_e})_{\mbox{data}} \over (N_{\nu_\mu} / N_{\nu_e})_{\mbox{MC}}}, 
\eeq
where the subscript `MC' for the ratio in the denominator corresponds to 
expectations based on Monte Carlo numerical simulations. 
Both IMB and Kamiokande have found this double ratio, $R$ to be of the 
order of 0.6 instead of 1 as one would have expected. Neutrino oscillations 
were again thought to be the culprit for this discrepancy. In 1998, the 
Super-Kamionkande collaboration announced strong evidence for neutrino 
oscillations in atmospheric neutrinos with high statistics. This was one
of the first evidences of neutrino oscillations with such experimental
accuracy and high statistics. These experiments observed an `up-down' 
asymmetry away from zero by about 10 standard deviations, putting the
phenomena of neutrino oscillations on firm experimental footing\footnote{The
up-down asymmetry is expected to be zero if there are no 
oscillations.}\cite{grimus}. 

Soudan-2 and MACRO experiments, both of which are based on iron calorimeters 
have further confirmed the hypothesis that atmospheric neutrinos do oscillate
and hence 
removing any suspicions regarding this phenomena being observed only at 
water Cerenkov detectors, perhaps due to some systematic errors particular
to those detectors. In the recent years, 
two experiments K2K and MINOS have further reduced the errors in the 
measurement of the oscillation parameters associated with the atmospheric 
neutrinos. They are now given to be as\cite{garcia} :
\begin{eqnarray}
\Delta m^{2}_{\mbox{atm}} &=& 2.6 \pm 0.2(0.6) \times 10^{-3}
\mbox{eV}^{2} \nonumber \\
\theta_{\mbox{atm}} &= & 43.3 ^{+4.3}_{-3.8} (^{+9.8}_{-8.8}) \deg,
\end{eqnarray}
where as before we have quoted the 1$\sigma$ (3$\sigma$) error bars.

Given these numbers for the mass squared differences and the mixing angles, 
we are now ready to reconstruct from the experimental data the neutrino
mass matrix \cite{resources}. As mentioned in the introduction, the Standard Model of
particle physics has made no provisions for non-zero neutrino masses. 
To accommodate for non-zero neutrino masses, several extensions of the
Standard Model have been considered. Experimentally, however
a few issues still need to be settled. These are (i) whether neutrinos 
are of Majorana nature or Dirac nature. This determines whether neutrinos
are anti-particles of themselves or not. This important issue could be
tested in future neutrinoless double beta decay experiments whose transitions
are only possible if neutrinos are  Majorana (self anti-particles) in nature.
This also has implications for the structure of the neutrino mass matrix 
as, in the Majorana case, the mass matrix is complex symmetric whereas in 
the Dirac case its complex generic. 
(ii) The second issue is related to the point that we have so far measured 
only the mass squared differences of the neutrinos but, not their absolute 
masses. With three neutrinos, we can have the observed mass squared differences
in three different hierarchies (a) Normal Hierarchy (NH) 
$m_{\nu_1}~ \ll m_{\nu_2}~ \ll m_{\nu_3}$; (b) Inverted Hierarchy (IH)
$m_{\nu_3}~ \ll m_{\nu_1}~ \ll m_{\nu_2}$; (c) Degenerate 
$m_{\nu_1}~ \sim m_{\nu_2}~ \sim m_{\nu_3}$. Future experiments based
on cosmology, long base line neutrino propagation and perhaps even neutrino less
double beta decay are expected to shed light on this important aspect of
the neutrino mass hierarchy. (iii) We have not yet measured 
the third neutrino mixing angle $\theta_{13}$ which appears in the three
neutrino mixing scheme. At present there is only an upper bound  from the
CHOOZ experiment in France and its present limits are given as 
$\theta_{13} = 0^{+5.2}_{-0.0} (^{+11.5}_{-0.0}) \deg$. Future experiments
like Double CHOOZ in France and Daya Bay in China are expected to improve
this limit by at least an order of magnitude. (iv) Finally we have the question
whether CP (a product of Charge conjugation symmetry and Parity )\footnote{This
symmetry plays an important role in our understanding of the origins of matter
and anti-matter asymmetry in our world.} is a good symmetry or not in the leptonic
sector. Experimentally this question is quite challenging and it crucially depends
on the value of unknown  neutrino mixing angle $\theta_{13}$. Future experiments
will hopefully able to uncover this mystery.

One of the most popular and simplest extensions of the Standard Model
gives neutrino masses through the so-called \textit{see-saw} mechanism. 
In this mechanism right handed neutrinos are added to the standard model
particle spectrum. Given that these particles do not obey Standard Model
symmetries, they can have very large masses  however breaking 
lepton number. At the same time, they can couple with the Standard Model
(left handed) neutrinos resulting in a lepton number conserving (Dirac)
mass, which  can be expected to be close to one of the masses of the other 
Standard Model particles like top quark, bottom quark or tau lepton etc. 
The interplay between the large Majorana mass and the Dirac mass leads to
a small non-vanishing mass $\sim $ eV to the SM left handed neutrinos, just
as what is expected by the experiments. 
It would be instructive to see what the structure of the neutrino mass 
matrix is as demanded by the data. In the below, we will assume that 
neutrinos are Majorana in nature (as 
indicated by the seesaw mechanism) and further follow normal hierarchy (NH).
In such a scheme, the neutrino mass matrix is given by :
\beq
\mathcal{M}_{\nu} = U_{\mbox{PMNS}}^\star \mathcal{M}_{\mbox{diag}} U_{\mbox{PMNS}}^{\dagger},
\eeq
where $\mathcal{M}_{\mbox{diag}} = 
\mbox{Diag}\{m_{\nu_1},m_{\nu_2},m_{\nu_3}\}$, with 
$m_{\nu_1} \ll \sqrt{\Delta m_{\mbox{solar}}^2}$, 
$m_{\nu_2} \sim \sqrt{\Delta m_{\mbox{solar}}^2}$, 
$m_{\nu_3} \sim \sqrt{\Delta m_{\mbox{atm}}^2}$. Neglecting the phases, 
the $U_{\mbox{PMNS}}$  has the form (at 3$\sigma$ level) given
 by \cite{garcia} :
\beq
U_{\mbox{PMNS}} = \left( \begin{array}{ccc}
0.79-0.86 & 0.50-0.61 & 0.0-0.20\\
0.25-0.53 & 0.47-0.73 & 0.56-0.79 \\
0.21-0.51 & 0.42-0.69 & 0.61-0.83 \end{array} \right)
\eeq 
Considering  the values for the individual neutrino masses depending on the scheme,
one can reconstruct the neutrino mass matrix. 
This summarises the present status of three neutrino mixing and oscillations
as we understand now. 

\section{LSND and KARMEN : Indications for a sterile neutrino}

While the search for a robust signal in solar and atmospheric neutrino 
oscillations was going on, simultaneously experimentalists have been on the 
look out for neutrino oscillations at other frequencies (ie, at $\Delta m^2$ 
other than those relevant for solar and atmospheric neutrino oscillations). 
Most of these earlier experiments had short base lines, typically about few
tens of meters\footnote{To probe solar and atmospheric neutrino oscillations on earth, 
one would need much larger base lines.} and 
are thus sensitive to $\Delta m^2 \ger 1 \mbox{eV}^2$. The LSND was
one such experiment.  Another important characteristic of the LSND experiment 
was that it was an appearance experiment. Typically, we can think of two types
of strategies while looking for neutrino oscillations :\\
(a) Disappearance experiments: Here, we look for a reduction in the expected
number of the neutrinos (which are detected) of a particular flavour. Then 
this disappearance is explained in terms of neutrino oscillations (in to 
undetected flavours). \\
(b) Appearance experiments: In other case, we can look for neutrino flavours 
which are either not present or very weakly produced at the neutrino source.  
Again this appearance is explained in terms of neutrino oscillations. 
It should be noted that earlier short based lined experiments
have not found any evidence for neutrino oscillations. The initial indications 
for oscillations in both solar and atmospheric sectors have come from various
disappearance experiments. 

The LSND was based at Los Alamos National Laboratory (LANL) in United States. LSND, 
which stands for Liquid Scintillation Neutrino Detector had a base line of 
30 Meters and was looking for an excess of $\nu_e,~ \bar{\nu}_e$,  
starting from a beam which was mainly made up of $\nu_\mu$ (and $\bar{\nu}_\mu$). 
The LSND has collected data from 1993 up to 1998, and the collaboration first 
reported `evidence' for anti-neutrino oscillations in 1995, thus becoming the
first experiment to report observation of neutrino oscillations using appearance
type strategy. 

The experimental set up is quite simple\cite{lsnd1}. The source of neutrinos was an intense 
proton beam at the Los Alamos Meson Physics Facility (LAMPF) whose kinetic energy 
is 800 MeV (~1 m A current). This beam was made to hit a water target,
followed by a water-cooled Copper (Cu) beam dump. This produces large numbers of
pions, mostly $\pi^+$. The $\pi^+$ decays in to a $\mu^+$ and $\nu_\mu$ and 
$\mu^+ \to e^+ \nu_e \bar{\nu}_\mu$. Not many electron anti-neutrinos ($\bar{\nu}_e$) 
are expected from such a source (small amounts of $\pi^-$ are produced, but are 
immediately absorbed, a few of them decay to $\mu^-$, which are also absorbed 
before decaying). As mentioned earlier, the LSND detector itself was 
situated about 30 Meters from the
source.  The detector is approximately a cylindrical tank 8.3 Meters long and of 
5.7 Meters diameter. It contained 167 tonnes of mineral oil (CH$_{2}$) and 
0.031 g/litre of b-PBD (butyl-phenyl-bipheny-oxydiazole) which acted as the 
organic scintillating medium filling the detector. The detector is lined up with
phototubes ($1220$ in number, 8-inch in size and of Hamamatsu make ) inside 
the tank to detect the
Cherenkov radiation as well as the scintillation light emitted from the propagating
particle inside the detector. Further the detector was adequately shielded from 
cosmic rays by an overburden of roughly 2 $\mbox{Kg}/\mbox{cm}^2$. 

Data was collected in two batches from 1993 to 1995 using the water target in
the neutrino source described above and later replacing the water target with
a closely packed high atomic number element (Z) from 1996 to 1998. Data from 
two types of decay patterns of  $\mu$ons was collected (i) $\mu$ decay at rest :
used for the analysis of anti-neutrinos (ii) $\mu$ decay in flight : used
for the analysis for neutrinos.  A total of 18 $\times ~10^{22}$ (\textit{i.e,} 
a trillion-billion, 180,000,000,000,000,000,000,000) protons were made to hit 
the LSND target during this period.  The following reactions were used to detect 
the $\overline{\nu_{e}}$ emanating from $\mu$ decays at rest: 
\beq
\overline{\nu_{e}} + p \rightarrow e^{+} + n
\eeq
and the 2.2 MeV $\gamma$ from the reaction
\beq
 n + p \rightarrow d + \gamma.
\eeq
In the case of $\mu$ decays in flight, the experiment looked for electron 
neutrinos which are expected to be present after oscillations of the muon 
neutrinos during the flight. The reaction used to detect the electron 
neutrino was
\beq
\nu_{e} + ^{12}C \rightarrow e^{-} + X
\eeq
the signal being the single electron, where $X$ stands for the residue 
of the $^{12}C$ atom due to this inelastic scattering.  

In the data analysis, the energy range was taken to be 20 $<$ E$_{e}$ $<$ 60 MeV 
for the $\overline{\nu_{\mu}} \rightarrow \overline{\nu_{e}}$ oscillation search and 
60 $<$ E$_{e}$ $<$ 200 MeV for the $\nu_{\mu} \rightarrow \nu_{e}$ oscillation search. 
In the anti-neutrino oscillation search,  a total excess of 
87.9 $\pm$ 22.4 $\pm$ 6.0 events(3.8 $\sigma$) consistent with 
$\overline{\nu_{e}} + p \rightarrow e^{+} + n$ scattering was observed above the 
 background. This excess corresponds to an oscillation probability of 
(0.264 $\pm$ 0.067 $\pm$ 0.045) percent assuming the the two anti-neutrino oscillation 
hypothesis. The neutrino oscillation search, in addition to the anti-neutrino
search also found an excess of events though statistically, this excess was 
not significant. It amounted to  8.1 $\pm~12.2~\pm~1.7$ events corresponding
to an oscillation probability of (0.10$\pm ~0.16~\pm~0.04)\%$. To summarise,  
the LSND data suggested that (anti)neutrino 
oscillation occurred with a $\Delta m^{2}$ in the range of 0.2 to 10 eV$^{2}$/c$^{4}$.
At 90 percent C.L. analysis of the $\mu^{+}$ decay at rest data showed that
sin$^{2}$ 2$\theta$ $\in$ [10$^{-3}$ to 10$^{-1}$].

The implications of the LSND result are many fold : firstly, it indicates
that the standard three flavour picture which we have summarised in the 
previous section would not longer hold true as with three neutrinos, one
can have only two independent mass squared differences. This can be easily
seen as follows ; the mass squared differences $\Delta m^2_{ab}$ as defined
below eq.(\ref{prob}), satisfy the following equation in three generations :
$\Delta m^2_{21} + ~\Delta m^2_{32} +~ \Delta m^2_{13} = 0$, which shows 
there are only two independent mass squared differences in three generations.
Secondly, if there
is another neutrino responsible for the oscillations observed at LSND, this
neutrino cannot be a part of the Standard Model families,
as it would violate the experimental result from the LEP experiment at CERN,
which said that three are only three families of neutrinos which take part
in the Standard Model (more precisely weak) interactions.  
Thus the new neutrino has to be a  \textit{inert} under these interactions
and thus named as a \textit{sterile} neutrino.

Theoretically, the existence of a sterile neutrino would require deeper
understanding of such particles\cite{resources}. Further newer mechanisms 
might be required
 to generate masses to them which can sometimes lead to complicated model 
building beyond the Standard Model. Phenomenologically too, simplest 
extensions from the three neutrino
scheme to the four neutrino scheme, including a sterile neutrinos to accommodate
the LSND data have run in to rough weather with improving measurements of
solar and atmospheric data which have serious implications on such schemes. 
This is because little room is left to accommodate a sterile neutrino either in the
solar data or in the atmospheric data. 
Finally the sterile neutrino can only be tested indirectly. Indications can
come from neutrino oscillations and perhaps through cosmology where sterile
neutrinos can play a role in structure formation. 
Sterile neutrinos also have severe constraints from astrophysical processes 
like supernovae cooling etc\cite{vissanistrumia}.
While all these would pose new exciting challenges, the existence of  a 
sterile neutrino experimentally relied only on the LSND data. 

The sterile neutrinos are not the only solution offered to understand LSND
data. Several new exotic ideas as well as some  well motivated theories 
were  used to explain the LSND data. For example, within supersymmetric 
extensions of the standard model new kinds of interactions which violate 
lepton number can be used to explain the LSND excess events. On the other
hand, well motivated models based on theories of extra space dimensions 
also have a natural way of incorporating sterile neutrinos and LSND data\cite{dudas}.  
In addition to these, more exotic ideas like 
CPT violation\cite{barenboimmurayama}, which 
advocates different masses for particles and anti-particles and ideas of mass varying
neutrinos which propose neutrino masses vary with time over cosmological time
scales have been put to use explain to the LSND data in the
recent years. 

\subsection{The KARMEN experiment}

The LSND result ran in to controversy when some experimentalists have 
raised objections on the estimation of systematical errors of the experiment. 
The LSND collaboration has responded to these concerns by changing the
target (from water to a closely packed high Z target) and further explaining
that there could not be large errors introduced in to the systematics due
to the presence of other sources of electron anti-neutrinos in the experiment. The KARMEN
experiment, which was studying neutrino-nucleus cross sections around that
time  was expected to 
provide an independent confirmation or verification of the LSND observations
after some modifications to their existing  experimental set up. 

This experiment whose acronym reads  KARMEN (KArlsruhe Rutherford Medium Energy Neutrino) 
was located 
at the highly pulsed spallation neutron source ISIS of the Rutherford 
Laboratory (UK). The experiment was most sensitive to the search of $\bar{\nu_\mu} \to
\bar{\nu_e}$ oscillation channel.

In this case, a rapid cycle synchrotron is used to accelerate the protons
upto 800 MeV  with a design beam current of 200 $\mu$A. The protons are 
made to hit a target of water-cooled Ta-D$_{2}$O, which produced $\pi^+$, which
decays then in to $\mu^+$; the subsequent decays of $\mu^+$ act as a source of
anti-muon neutrinos.
The detector which is a segmented high resolution liquid scintillation calorimeter,
is located at a mean distance of 17.7 Meters from the target.
The liquid scintillator consists of a mixture of paraffin oil (75 percent by volume), 
pseudocumene (25 percent by volume) and 2 g l$^{-1}$ of the scintillating active 
1-phenyl-3-mesityl-2-pyrazoline(PMP).  Appearance of $\overline{\nu_{e}}$ from 
$\overline{\nu_{\mu}} \rightarrow \overline{\nu_{e}}$ flavour oscillation.
is detected by the classical inverse beta-decay reaction:
\begin{eqnarray}
\overline{\nu_{e}} + p &\rightarrow& n + e^{+} ~~~ Q = -1.804 \mbox{MeV} \nonumber \\
n_{th} + ^{1}H &\rightarrow& ^{2}H + \gamma \nonumber \\
n_{th} + Gd &\rightarrow& Gd + n\gamma, 
\end{eqnarray}
where, the average number of photons emitted , $<n>$ = 3.
In total 15 candidates fulfilled all conditions for the 
$\overline{\nu_{e}}$ signature. This agreed with the
background expectation of 15.8 $\pm$ 0.5 events. Hence there was no 
signature of oscillations. Analysis of the data yielded the following results:
sin$^{2}2\theta < 1.7 \times 10^{-3}$ for $\Delta m^{2} \geq 100$~ eV$^{2}$
and $\Delta m^{2} < 0.055$~ eV$^{2}$ for sin$^{2}2\theta =1$
at 90 \% CL.  The implications are that at large $\Delta m^{2}$, KARMEN results exclude 
the region favoured by LSND. At low $\Delta m^{2}$ there is a restricted 
parameter region statistically compatible with both the experimental results.
A joint analysis with LSND shows that these results are 64 percent compatible
with each other\cite{combinedlsndkarmen}.

\section{MiniBooNE}

In order to address the LSND anomaly the MiniBooNE (BooNE is an acronym for 
the Booster Neutrino Experiment) experiment was proposed. The MiniBooNE 
collaborators have kept the $L/E$ same as in LSND but have changed
the systematics, energy and the event signature. This way, one has access to 
the entire parameter space accessed by LSND. 

MiniBooNE is located at the Fermi National Accelerator Laboratory,
Batavia, IL, in the United States. The experiment made use of
the Fermilab Booster neutrino beam. Protons with energies of 8 GeV
were incident on a Beryllium (Be) target; such a choice of the target
solely being dictated by the need of a source with far more $\mu^+$'s
than $\mu^-$'s. To increase the flux, a magnetic focusing horn which
encloses the target has been used (this increases the flux almost six
fold).  A total 6.3 $\times 10^{12}$ POT were delivered, while the
actual result of the experiment corresponded to
(5.58 $\pm$ 0.12)$\times 10^{12}$ POT\cite{miniboonerefs}.

This experiment was also based on the `appearance' principle ; it had
looked for an excess of $\nu_e$ in a purely $\nu_\mu$ beam. After the
protons hit the target, the produced (positively charged) pions and
kaons pass through a collimator of about 60 cm long and then through
a tunnel towards the detector which is about 50 m long. These particles
decay along the way producing neutrinos. The `intrinsic'
$\nu_{e} + \overline{\nu_{e}}$ sources are:
$\mu^{+} \rightarrow e^{+} + \nu_{e} + \overline{\nu_{\mu}}$ (52 percent)
$K^{+} \rightarrow \pi^{0} + e^{+} + \nu_{e}$ (29 percent)
$K^{0} \rightarrow \pi + e + \nu_{e}$ (14 percent)
others (5 percent) ;
$\nu_{e}/\nu_{\mu}$ = 0.5 percent and the anti-neutrino content
is about 6 percent.

The detector is placed about 541 m downstream in front of the target.
It has the shape of a sphere, 12.2 m in diameter.
This spherical tank is filled up with 800 tonnes of pure mineral oil
(fiducial volume = 450 tonnes \footnote{Actual volume relevant in the
detection process.}). An optical barrier separates the detector
into 2 regions (a) an inner light-tight volume of radius 575 cm and
(b) an optically isolated outer volume 35 cm thick known as veto region.
The optical barrier is lined with 1280 inner photomultiplier tubes (PMT)
(8-inch) providing 10 percent photocathode coverage. An additional
240 veto phototubes are lined in the inner volume detecting particles
entering or leaving the detector.

The produced neutrinos traverse along the tunnel, enter the detector,
and interact with the medium in the detector.  Depending on the pattern
of light observed in the PMTs, one can determine the kind of interaction
the neutrino went through in the detector. Two signatures (a) Cherenkov 
radiation and (b) scintillation (fluorescence) light are used to detect 
the kind of neutrino interaction. Neutrinos interact through both charged 
current and neutral current channels here and both are used in the detection
process. The main interactions are (1) Charged-current scattering (39 percent), 
(2) Neutral current(NC) elastic scattering (16 percent), (3) Charged-current 
(CC) single pion production (29 percent) (4) NC single pion production (12 percent), 
(5) Multi-pion and deep-inelastic scattering (less than 5 percent).
The list of all possible interactions and the corresponding signature in the 
PMT can be found in the research paper put out by the collaboration\cite{miniboonerefs}.
For example, in the charged- current quasi-elastic events, a neutrino
interaction in the detector will produce the lepton partner of
the neutrino. Electrons multiple-scatter along their way and so travel for
a very short time before their velocity falls below that required for
Cherenkov radiation. Hence a fuzzy Cherenkov ring in the detector is their
signature. Muons, being heavier, have much longer tracks. As they slow down,
the angle at which the Cherenkov light is being emitted shrinks. Muons also
emit scintillation light. The signature is a sharp outer ring with fuzzy
inner region.Neutral pions decay into 2 photons which then pair-produce
(an electron and a positron). Evidently their signature in the detector
are two fuzzy rings.

Data was collected for about five years starting from 2002. After the data was taken, 
the MiniBooNE collaboration performed a "blind" analysis. This means the experimentalists 
did not have access to all the information in the data. This is one of the hallmarks
of the work done by this collaboration. For oscillation search two different 
types of analysis were performed : one which depended on likelihood variables (called 
the "Track Based" ,TB analysis), and the one which depended on a boosted decision tree. In this
way, each analysis would cross-check the other analysis. In the published analysis, the 
former algorithm was chosen as the primary result because it had a better sensitivity 
to $\nu_{\mu} \rightarrow \nu_{e}$ oscillation.
In the analysis,the electron neutrino events were isolated and then  a comparison 
is made between the observed number of events to the expected number of events 
(that is the sum of the intrinsic electron neutrino and the fake events) as a 
function of the `reconstructed' neutrino energy. An excess of the observed data 
over expected data (or an excess of $\nu_{e}$ events) as a function of the energy 
indicates oscillation.

After the complete analysis was done "the box" was opened : it was found that 
there was no significant excess of events (22 $\pm$ 19 $\pm$ 35 events) for 
475 $<$ E$^{QE}_{\nu} <$ 1250 MeV.  The oscillation fit in the 
475 $<$ E$^{QE}_{\nu} <$ 1250 MeV range yields a$\chi^{2}$ probability of 
93 percent for the null hypothesis, and a probability of 99 percent for the 
(sin$^{2} \theta$ = 10$^{-3}$, $\Delta m^{2}$ = 4 eV$^{2}$) for 
the best-fit point. The probability that MiniBooNE and LSND both are due 
to two-neutrino oscillations is only 2 percent\footnote{This result is
only true as long as one restricts $\Delta m^2$ from  $0.2$ to $2.75$ 
eV$^2$\cite{heather}.  A recent analysis by MiniBooNE collaboration
finds that there is a non-negligible probability that the results from
all the three experiments, namely, LSND, Karmen and Mini BooNE are due 
to a two-neutrino oscillation if the $\Delta~ m^2$ is taken to be much
lower\cite{heather}.}.

\section{Implications of MiniBooNE and Future Directions}

The MiniBooNE's results will have strong implications for most of the sterile neutrino 
models which are constructed as extensions of the Standard Model. However, before that
the MiniBooNE still has some things which have to be understood about its own analysis. 
The experiment has reported an excess of events (96 $\pm$ 17 $\pm$ 20 events) 
( deviation = 3.7 $\sigma$ ) was observed below 475 MeV above the expected background.
Presently, very little understanding is present about the source of this excess. It is
not clear whether it is an experimental systematical error or whether it signals the 
existence of new physics. 

One of the major implications of the MiniBooNE result is that simplest sterile 
neutrino schemes, like $3+1$ or $2+2$ with single sterile neutrino are ruled out
as they are not compatible with both LSND and MiniBooNE data. However, the exploiting
CP violation present in much larger schemes like $3+2$ with two sterile neutrinos 
can still accommodate LSND and MiniBooNE data making them compatible\cite{sruba}. 
The case of this larger $3+2$ framework has implications also for astrophysical
neutrinos, especially neutrinos involved in supernovae\cite{sandhya1}. It is 
also proposed that sterile neutrino signatures can be found at neutrino telescopes
probing ultra high energy neutrinos\cite{sandhya2}.

	Mass varying neutrinos have been proposed as means of generating
	the cosmological dark energy in the recent years. Here the neutrinos
	have couplings to an \textit{acceleron} field which vary over
	cosmological times scales. This idea has been applied to explain
	the LSND data. Just as in the three neutrino case, here too one
	would need to add another neutrino to accommodate the LSND data 
	as we would need at least one more mass squared difference in
	addition to the ones required. It has been pointed out that 
	in this particular model\cite{bargermarfy1}, 
	it could happen that there could be positive signal at LSND 
	whereas a null result for MiniBoone. 
	How far this idea would remain viable with future long based
	experiments remains to be seen. 
	 
	While CPT violation need is not completely understood within
	the context of quantum field theory, in the neutrino sector
	it can incorporated by assuming neutrinos and anti-neutrinos
	have different masses and mixing angles and thus the oscillation
	frequencies of neutrinos and anti-neutrinos would be different.
	This has been utilised to explain the LSND data. However after
	the KamLand experiment, there has been some skepticism though
	it was shown that statistically the fits could be still reasonable. 
	The fate of a four\cite{bargermarfy2} or high number of 
	neutrino generation CPT violating models needs to be seen.

Thus, at present the last word has not yet been said about the fascinating world
of sterile neutrinos. As MiniBooNE continues to take data, we expect more severe
constraints from them. 

\noindent
\textbf{Acknowledgments}
We thank R. Godbole for encouraging us to write this article. Thanks are also due
to B. Ananthanarayan for his wonderful talk on this subject and carefully reading 
the manuscript.


\begin{thebibliography}{99}
\bibitem{pontecorvo}
B.~Pontecorvo,
``Mesonium and antimesonium,''
Sov.\ Phys.\ JETP {\bf 6}, 429 (1957)
[Zh.\ Eksp.\ Teor.\ Fiz.\  {\bf 33}, 549 (1957)] ;\\
%
S.~M.~Bilenky and B.~Pontecorvo,
``Lepton Mixing And Neutrino Oscillations,''
Phys.\ Rept.\  {\bf 41}, 225 (1978).
%
\bibitem{grimus}
S.~M.~Bilenky, C.~Giunti and W.~Grimus,
``Phenomenology of neutrino oscillations,''
Prog.\ Part.\ Nucl.\ Phys.\  {\bf 43}, 1 (1999)
[arXiv:hep-ph/9812360].
%
\bibitem{ananth}
See for example, articles of 
B.~Ananthanarayan and R.~K.~Singh,
``Direct observation of neutrino oscillations at the Sudbury Neutrino
Observatory,''
Curr.\ Sci.\  {\bf 83}, 553 (2002)
[Resonance J.\ Sci.\ Educ.\  {\bf 7N10}, 79 (2002\ ERRAT,8N1,88.2003)]
[arXiv:physics/0208096] ;\\ 
B.~Ananthanarayan, C.~Gautham, A.~Mavalankar, K.~Shivaraj, S.~Uma Sankar and A.~Upadhyay,
``Research news: Progess in determination of neutrino oscillation
parameters,''
Curr.\ Sci.\  {\bf 91}, 864 (2006)
[arXiv:physics/0608152] and references there in.
%
\bibitem{garcia}
For a recent reviews see, 
  S.~Goswami, A.~Bandyopadhyay and S.~Choubey,
  ``Global analysis of neutrino oscillation,''
  Nucl.\ Phys.\ Proc.\ Suppl.\  {\bf 143}, 121 (2005)
  [arXiv:hep-ph/0409224] ;\\
%
M.~C.~Gonzalez-Garcia and M.~Maltoni,
``Phenomenology with Massive Neutrinos,''
arXiv:0704.1800 [hep-ph].
\bibitem{resources}
See for example, the following resource papers :
R.~N.~Mohapatra {\it et al.},
``Theory of neutrinos,''
arXiv:hep-ph/0412099 ;\\
%
R.~N.~Mohapatra {\it et al.},
``Theory of neutrinos: A white paper,''
arXiv:hep-ph/0510213.
%
\bibitem{lsnd1}
C.~Athanassopoulos {\it et al.}  [LSND Collaboration],
``Evidence for anti-nu/mu --> anti-nu/e oscillation from the LSND  experiment
at the Los Alamos Meson Physics Facility,''
Phys.\ Rev.\ Lett.\  {\bf 77}, 3082 (1996)
[arXiv:nucl-ex/9605003] ;\\
%
A.~Aguilar {\it et al.}  [LSND Collaboration],
``Evidence for neutrino oscillations from the observation of anti-nu/e
appearance in a anti-nu/mu beam,''
Phys.\ Rev.\  D {\bf 64}, 112007 (2001)
[arXiv:hep-ex/0104049].
%
\bibitem{vissanistrumia}
For a nice review on sterile neutrinos, please see
M.~Cirelli, G.~Marandella, A.~Strumia and F.~Vissani,
``Probing oscillations into sterile neutrinos with cosmology,  astrophysics
and experiments,''
Nucl.\ Phys.\  B {\bf 708}, 215 (2005)
[arXiv:hep-ph/0403158].
%
\bibitem{dudas}
K.~R.~Dienes, E.~Dudas and T.~Gherghetta,
``Light neutrinos without heavy mass scales: A higher-dimensional seesaw
mechanism,''
Nucl.\ Phys.\  B {\bf 557}, 25 (1999)
[arXiv:hep-ph/9811428].
%
\bibitem{barenboimmurayama}

G.~Barenboim, L.~Borissov, J.~D.~Lykken and A.~Y.~Smirnov,
``Neutrinos as the messengers of CPT violation,''
JHEP {\bf 0210}, 001 (2002)
[arXiv:hep-ph/0108199] ;\\
%
H.~Murayama and T.~Yanagida,
``LSND, SN1987A, and CPT violation,''
Phys.\ Lett.\  B {\bf 520}, 263 (2001)
[arXiv:hep-ph/0010178].
%
\bibitem{combinedlsndkarmen}
B.~Armbruster {\it et al.}  [KARMEN Collaboration],
``Upper limits for neutrino oscillations anti-nu/mu --> anti-nu/e from  muon
decay at rest,''
Phys.\ Rev.\  D {\bf 65}, 112001 (2002)
[arXiv:hep-ex/0203021].
%
\bibitem{miniboonerefs}
A.~A.~Aguilar-Arevalo {\it et al.}  [The MiniBooNE Collaboration],
``A Search for Electron Neutrino Appearance at the Delta m**2 ~ 1 eV**2
Scale,''
Phys.\ Rev.\ Lett.\  {\bf 98}, 231801 (2007)
[arXiv:0704.1500 [hep-ex]];
%
See also, the website :
http://www-boone.fnal.gov/ and links there in.
%
\bibitem{heather}
Heather Ray, Mini BooNE collaboration, by private communication.

\bibitem{sruba}
M.~Maltoni and T.~Schwetz,
``Sterile neutrino oscillations after first MiniBooNE results,''
arXiv:0705.0107 [hep-ph]; \\
%
S.~Goswami and W.~Rodejohann,
``MiniBooNE Results and Neutrino Schemes with 2 sterile Neutrinos:   Possible
Mass Orderings and Observables related to Neutrino Masses,''
arXiv:0706.1462 [hep-ph].
%
\bibitem{sandhya1}
  S.~Choubey, N.~P.~Harries and G.~G.~Ross,
  Phys.\ Rev.\  D {\bf 74}, 053010 (2006)
  [arXiv:hep-ph/0605255];
%
  S.~Choubey, N.~P.~Harries and G.~G.~Ross,
  Phys.\ Rev.\  D {\bf 76}, 073013 (2007)
  [arXiv:hep-ph/0703092].
%

\bibitem{sandhya2}
  R.~L.~Awasthi and S.~Choubey,
  arXiv:0706.0399 [hep-ph];
%
  S.~Choubey,
  arXiv:0709.1937 [hep-ph].
%

\bibitem{bargermarfy1}
V.~Barger, D.~Marfatia and K.~Whisnant,
``Confronting mass-varying neutrinos with MiniBooNE,''
Phys.\ Rev.\  D {\bf 73}, 013005 (2006)
[arXiv:hep-ph/0509163].
%
\bibitem{bargermarfy2}
V.~Barger, D.~Marfatia and K.~Whisnant,
``LSND anomaly from CPT violation in four-neutrino models,''
Phys.\ Lett.\  B {\bf 576}, 303 (2003)
[arXiv:hep-ph/0308299].
\end{thebibliography}
\end{document}